\documentclass[12pt,preprint]{aastex}
\usepackage{amsfonts}
\usepackage{gensymb}
\usepackage{enumerate}
\usepackage{subfigure}
\begin{document}
%
%
%
%TITLE
\title{Dusty WDs in the \emph{WISE} All Sky Survey $\cap$ SDSS\footnotemark[1]}
\author{
Sara D. Barber\altaffilmark{2,4},
Mukremin Kilic\altaffilmark{2},
Warren R. Brown\altaffilmark{3},
A. Gianninas\altaffilmark{2}
}
\altaffiltext{2}{Homer L. Dodge Department of Physics and Astronomy, University of Oklahoma, 440 W. Brooks St., Norman, OK 73019, USA}
\altaffiltext{3}{Smithsonian Astrophysical Observatory, 60 Garden Street, Cambridge, MA 02138, USA}
\altaffiltext{4}{barber@nhn.ou.edu}

\footnotetext[1]{Based on observations obtained at the MMT Observatory, a joint facility of the Smithsonian Institution and the University of Arizona.}
%
%
%
%ABSTRACT
\begin{abstract}
A recent cross-correlation between the SDSS DR7 White Dwarf Catalog with the Wide-Field Infrared Survey Explorer (\emph{WISE}) all-sky photometry at 3.4, 4.6, 12, and 22 microns performed by \cite{debes11} resulted in the discovery of 52 candidate dusty white dwarfs (WDs). The 6$\arcsec$ \emph{WISE} beam allows for the possibility that many of the excesses exhibited by these WDs may be due to contamination from a nearby source, however. We present MMT$+$SWIRC $J$- and $H$-band imaging observations (0.5-1.5$\arcsec$ PSF) of 16 of these candidate dusty WDs and confirm that four have spectral energy distributions (SEDs) consistent with a dusty disk and are not accompanied by a nearby source contaminant. The remaining 12 WDs have contaminated \emph{WISE} photometry and SEDs inconsistent with a dusty disk when the contaminating sources are not included in the photometry measurements. We find the frequency of disks around single WDs in the \emph{WISE} $\cap$ SDSS sample to be 2.6-4.1\%. One of the four new dusty WDs has a mass of $1.04~M_{\odot}$ (progenitor mass $5.4~M_{\odot}$) and its discovery offers the first confirmation that massive WDs (and their massive progenitor stars) host planetary systems.
\end{abstract}

%
%
%
%INTRODUCTION
\section{INTRODUCTION}
A new class of dusty white dwarfs (WDs) is rapidly being populated through infrared excess searches using \emph{Spitzer}, the Wide-Field Infrared Survey Explorer (\emph{WISE}; \citealt{wright10}), and various ground based telescopes such as the IRTF and Gemini (\citealt{zuckerman87,kilic05,becklin05,vonhippel07,jura07a,farihi10,debes11,barber12,xu12}). These WDs exhibit excess emission in the near- and mid-infrared due to the thermal reprocessing of light by a disk of circumstellar dust. These dust disks are the debris resultant from the tidal disruption of an asteroid that has veered off its orbit and passed within the WD's Roche lobe (\citealt{debes02, jura03,jura08,bonsor11,debes12}). The origin of these asteroids is dynamical instability, initiated by post-main sequence stellar mass loss, whence a massive planet (if present) will start to gravitationally interact with smaller planetary bodies. After numerous interactions, an asteroid's orbit will become increasingly eccentric and in some cases approach close enough to the WD to be ripped apart by tidal forces. The debris produced by this disruption embodies a circular disk geometry after many subsequent orbits (\citealt{debes12}).

Dust persists around a WD inside the tidal radius of the WD and outside the radius at which the equilibrium temperature is such that the dust will sublimate. Viscous torques cause the sublimated dust to accrete onto WD's surface (\citealt{rafikov11a,rafikov11b,veras13}). This accretion results in a spectroscopically detectable pollution of the otherwise pristine WD atmosphere. Photospheric abundance analyses of these WDs show that the accreted metals originate from tidally disrupted minor bodies similar in composition to that of bulk Earth (\citealt{zuckerman07,klein10,klein11,dufour10,dufour12,xu14}). Since at least one planet is required to perturb minor bodies out of their stable orbits (\citealt{debes12}), photospheric pollution as well as circumstellar debris disks serve as tracers for remnant planetary systems at WDs. 

Dusty disks surrounding WDs are generally assumed to be geometrically flat and optically thick (\citealt{jura03,rafikov11a,rafikov11b}). Based on flat disk models, they vary in width from a narrow ring, a few tenths of a solar radius, to a disk filling the entire region interior to the tidal radius of the star and exterior to the radius of dust sublimation. The WDs known to host debris disks typically range in temperature from 9500-24,000 $K$ and in mass from 0.5-0.9 $M_{\odot}$. With this work, we expand the parameter space occupied by WD+disk systems with the confirmation of the first disk orbiting a WD hotter than 24,000 $K$ and the first disk orbiting a WD more massive than 1 $M_{\odot}$.

We present follow up near-infrared (NIR) $J$- and $H$-band photometry using the 6.5 m MMT with SWIRC of 16 WDs exhibiting a mid-infrared excess indicative of a debris disk detected by \emph{WISE} (\citealt{debes11}). We use the higher spatial resolution of SWIRC to confirm the presence of a debris disk or to identify a photometric contamination in the six arcsecond \emph{WISE} beam. We start in Section \ref{target} with the selection of our targets. Section \ref{obs} is a description of the observations and data reduction as well as the debris disk modeling of the detected excesses. Finally, we present our results in Section \ref{results}.
%
%
%
%TARGET SELECTION
\section{TARGET SELECTION}\label{target}
The \emph{WISE} mission opens up new opportunities for detecting excess infrared light at WDs caused by orbiting planetary debris. \emph{WISE} provides four bands of mid-infrared photometry from 3.4 to 22 $\mu$m. The \emph{WISE} InfraRed Excesses around Degenerates (WIRED; \citealt{debes11}) Survey reveals a large number of candidate WD$+$dust disk systems through a cross-correlation of $\sim$18,000 WDs from the Sloan Digital Sky Survey (SDSS) Data Release 7 WD catalog (\citealt{kleinman13}), the 2MASS All Sky Data Release Point Source Catalog (\citealt{skrutskie06}), and the UKIRT Infrared Deep Sky Survey (UKIDSS) Data Release 5 Point Source Catalogs from the Large Area Survey and Galaxy Cluster Surveys (\citealt{lawrence07}). However, due to the large size of the \emph{WISE} beam, some of these detections are likely the result of nearby contaminating sources.

We use the SDSS and UKIDSS observations to pare down the sample of 52 disk candidates identified by the WIRED survey to a target sample of 24 WDs. We removed a total of 28 of the 52 candidate dusty WDs from our target list for various reasons; 
\begin{enumerate}[(1)]
	\item We found 10 to have a blending source in their SDSS images.\\
	085742.05+363526.6, 095337.97+493439.7, 100145.03+364257.3, 124455.15+040220.6,\\
	131641.73+122543.8, 134800.05+282355.1, 141351.95+353429.6, 151747.51+342209.7, \\
	155359.87+082131.3, 165012.47+112457.1
	\item Another seven have a blending source in their UKIDSS images.\\
	024602.66+002539.2, 082624.40+062827.6, 084303.98+275149.6, 092528.22+044952.4,\\
	133212.85+100435.2, 153149.04+025705.0, 224626.38$-$005909.2
	\item Six more are known in the literature to host circumstellar dust.\\
	030253.09$-$010833.7, 084539.17+225728.0, 104341.53+085558.2, 122859.93+104032.9,\\
	145806.53+293727.0, 161717.04+162022.3
	\item Two quasars were misidentified as WDs.\\
	031343.07$-$001623.3, 103757.04+035023.6 
	\item One final target is known to have a brown dwarf companion (\citealt{steele09}).\\
	222030.69$-$004107.3
	\item Although 011055.06+143922.2 and 090522.93+071519.1 do not have obvious blends in their SDSS and UKIDSS images, we did not target them because NIR SWIRC observations in addition to UKIDSS data would not offer new information about the SEDs of those WDs. These two WDs will reenter our discussion when we consider the frequency of disks in the WISE sample.
\end{enumerate}
These cuts reduce the number of true candidate disk systems in \emph{WISE} from 52 to 24 WDs without an obvious nearby blending source, not already known to host disks or brown dwarf companions, without NIR data, and correctly identified as WDs. Hereafter, the WDs in the WIRED sample will be referred to with the letter J followed by the first four numbers of their WIRED name.
%
%
%
%OBSERVATIONS AND DATA ANALYSIS
\section{OBSERVATIONS AND DATA ANALYSIS}\label{obs}
%
%Near-infrared Photometry
\subsection{Near-infrared Photometry}
During two nights in May 2012, we imaged 16 of our 24 targets in the NIR $J$- and $H$-band using the SAO Wide-Field InfraRed Camera (SWIRC; \citealt{brown08}) with the MMT located on Mount Hopkins near Tucson, Arizona. SWIRC has a 2048$\times$2048 pixel HAWAII-2 detector with a $0.15$ arcsec/pixel plate scale and $5.12 \times 5.12$ arcmin field of view. Exposure times were adjusted to approach a signal-to-noise of 10. Each target was observed for two sets of a 3$\times$5 dither pattern with offsets of 10 arcseconds to sample the sky background. We took dark frames for each exposure time and sky flats each evening. The weather conditions were good with an average seeing of 0.9 arcseconds. 

We used the SWIRC data reduction pipeline to dark-subtract and flat-field the images. The images were then shifted to a common position and mean combined into one final image for each target. We calculated a zeropoint offset for each combined image by taking a weighted mean of the offsets for the 2MASS stars in the field. We used the IRAF aperture photometry package $\verb=apphot=$ to measure the flux of each isolated WD and comparison 2MASS stars. In many cases, the presence of an additional point source very near to the target WD interfered with our aperture photometry measurements. We measured these blended targets using the IRAF $\verb=daophot=$ package, where we fit the point-spread-functions of the WD and nearest blending neighbor simultaneously.
%
%DiskModeling
\subsection{Disk Modeling}
We combine SDSS DR7 \emph{ugriz} (\citealt{kleinman13}) and \emph{WISE} 3.4 \& 4.6 $\mu$m photometry with our SWIRC $J$- \& $H$-band data to construct the SED of each WD. We model the flux of the WD in the NIR and MIR bands using pure hydrogen atmosphere models (\citealt{tremblay09,tremblay10,gianninas11}) which we normalize to the SDSS $griz$-band data. In most cases, a secondary point source appears within six arcseconds of the WD of interest. The SEDs of these WDs are shown in Figure \ref{blends} where the SDSS, SWIRC, and \emph{WISE} data appear as magenta circles, orange squares, and green diamonds, respectively. The SWIRC photometry of the blending point source (when present) is shown by empty orange squares. Two of the objects in this figure (J0914 and J1503) do not have obvious \emph{WISE} contaminants but their SEDs are not consistent with a disk and they likely have unresolved background contaminants.

\begin{figure}
	\centering
	\includegraphics[width=\linewidth]{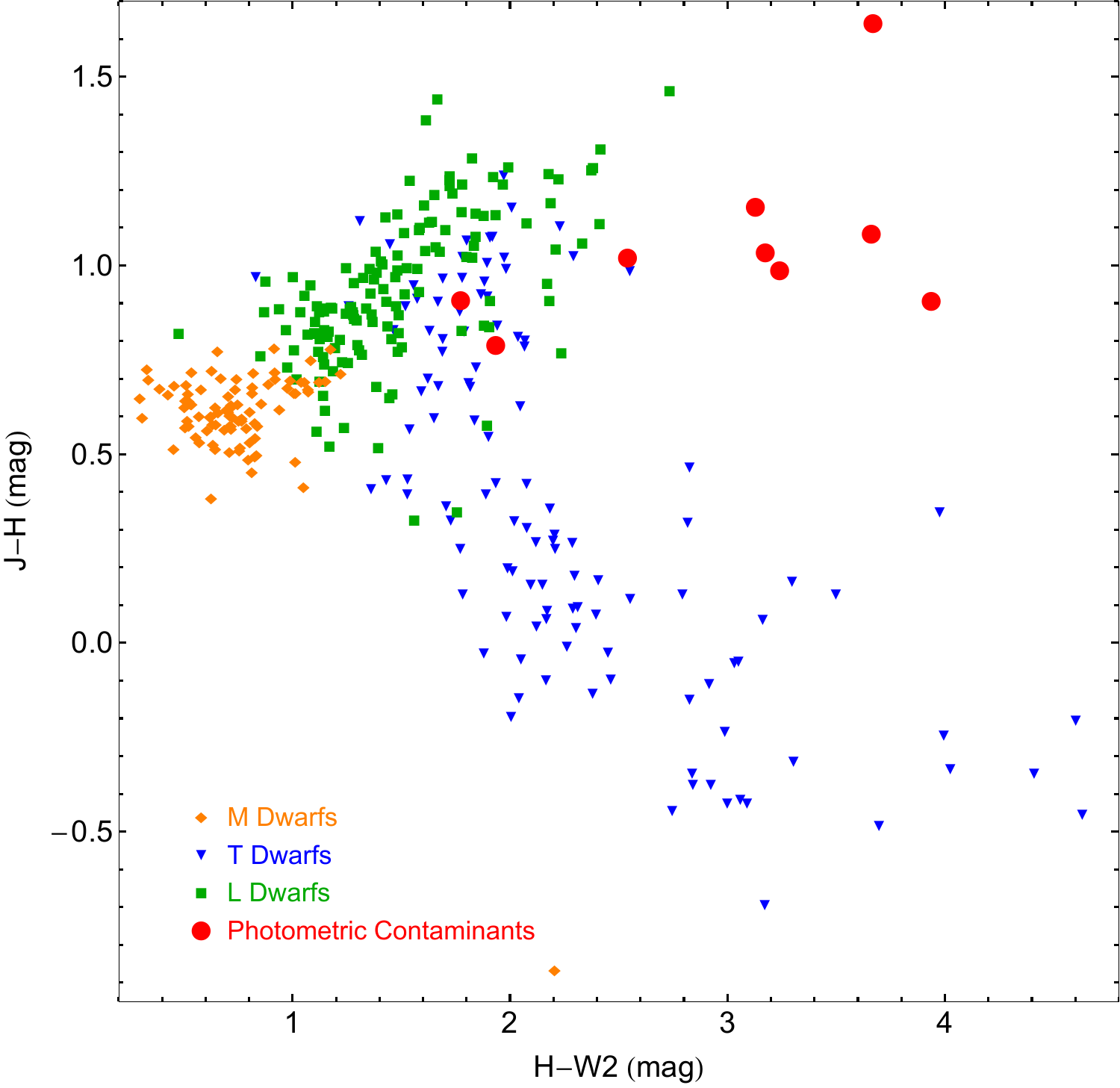}
	\caption{The $J-H$ and $H-W2$ colors of the contaminating sources near to the WDs in our sample are shown by red circles. The L/T dwarf and M dwarf colors are derived from the NIR and \emph{WISE} photometry published in Table 1 of \cite{kirkpatrick11} and are represented by green squares/blue triangles and orange diamonds, respectively. The colors of the contaminating sources in our SWIRC images are generally redder than the known brown dwarf and M dwarf populations and are more likely background galaxies.\label{blendcolors}}	
\end{figure}

The colors of the blending sources do not resemble those of M dwarf stars or L/T brown dwarfs (see Figure \ref{blendcolors}). The average $J-H$, and $H-W2$ colors of the blends in our images are 1.05 and 3.04, respectively, while the $J-H$ and $H-W2$ colors for typical T dwarfs are 0.36 and 2.25, for L dwarfs $J-H=0.92$ and $H-W2=1.49$ on average, and for typical M dwarfs $J-H=0.59$ and $H-W2=0.75$ (\citealt{west11,lepine11,kirkpatrick11,frith13,newton13}). Therefore, the blends in our sample are too red to be L/T brown dwarfs or M dwarf companions and are likely background galaxies.

In the cases where no obvious source of \emph{WISE} contamination is present (J1147, J1234, J1507, J1537), we rule out the possibility of not detecting the presence of a dim resolved contaminant. If the excesses at these four WDs were due to an resolved contaminant with the same $J-H$ and $H-W2$ colors as the blending sources described above, we should have detected them. The limiting magnitudes of our $J$- and $H$-band images are fainter than the expected brightnesses of the typical blending sources for the other WD targets.

\begin{figure}[h!]
	\centering
	\includegraphics[width=\linewidth]{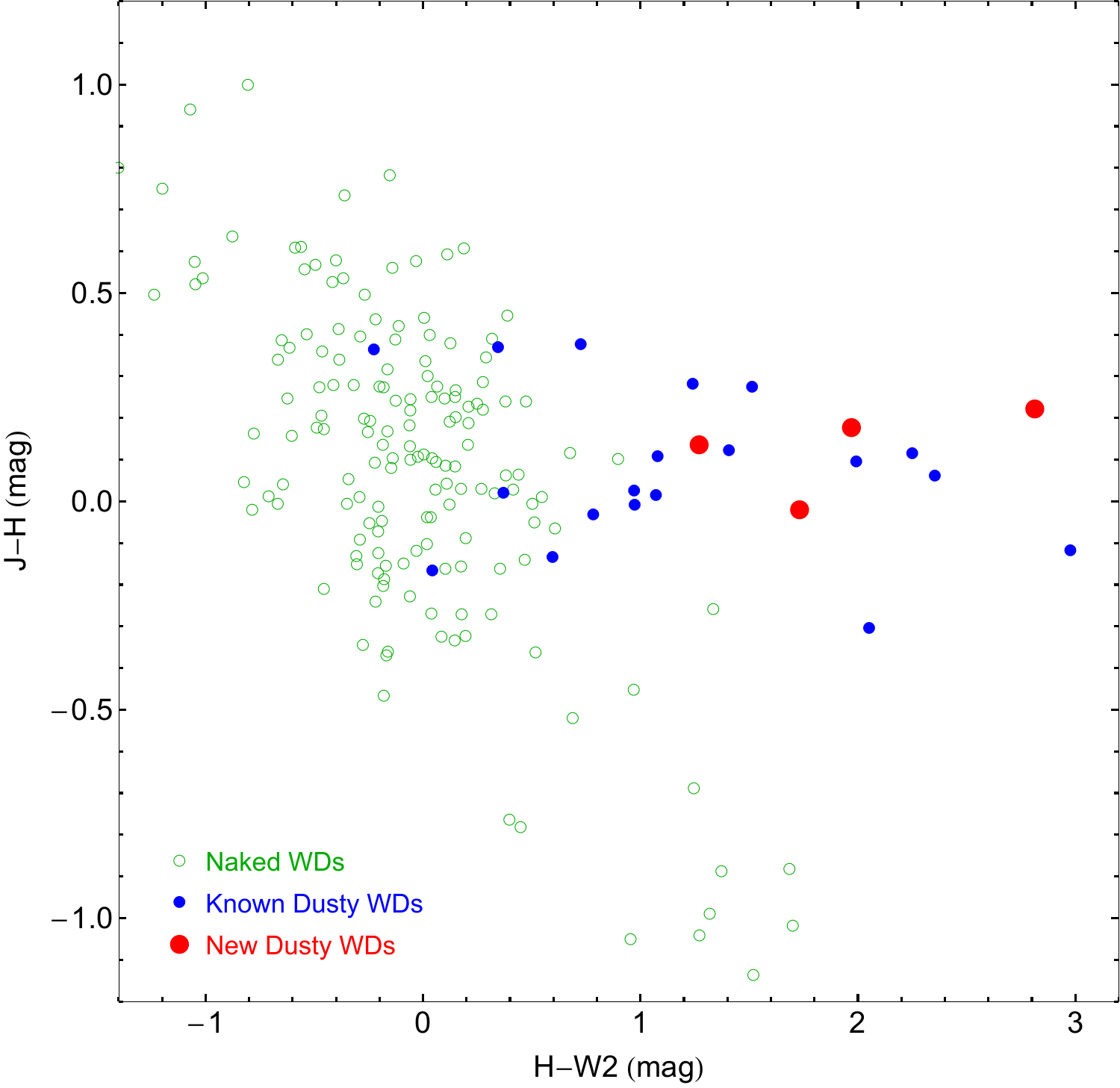}
	\caption{The $J-H$ and $H-W2$ colors of the naked WDs in the WIRED sample are shown as empty green circles, while the colors of known dusty WDs and our four new dusty WDs are shown as small blue and large red circles, respectively. The NIR and MIR photometry for the published WD$+$disk systems are taken from the 2MASS and \emph{WISE} online databases. The colors of the four new dusty WDs are comparable to the colors of the dusty WD population from the literature and are therefore likely to have the same mechanism generating their infrared excesses.\label{diskcolors}}
\end{figure}

For these four WDs, we fit a geometrically flat, optically thick disk model (\citealt{jura03}) to the SWIRC$+$\emph{WISE} excesses using a $\chi^2$ minimization method. Here $\chi^2$ is defined as the square of the difference between the measured excess and the disk model flux over the square of the photometric error for the SWIRC $J$- \& $H$- and the \emph{WISE} 3.4 \& 4.6 $\mu$m bands. We fit disk models with inner temperature of $800-2100~K$ and outer temperature $100-1200~K$ in steps of $10~K$ and inclination $0-90\degree$ in steps of $10\degree$. The SEDs are shown in Figure \ref{dusty} where the best fit disk model appears in red and the WD$+$disk model combination appears in black. Debris disk models accurately reproduce the excesses found at the four isolated WDs in the sample. We compare the colors of these new dusty WDs to the previously known dusty WDs from the literature in Figure \ref{diskcolors}. The colors of the four new dusty WDs are comparable to the colors of the dusty WD population from the literature and are therefore likely to have the same mechanism generating their infrared excesses. 
%
%
%
%RESULTS
\section{RESULTS}\label{results}
%
%Blended Targets
\subsection{Blended Targets}\label{blendedtargets}
The SWIRC photometry of our sample along with physical characteristics of the WDs are tabulated in Table \ref{sample}. The majority of our disk-candidate targets reveal extraneous sources of contamination when observed with high spatial resolution imaging in the NIR. We measured the photometry of the WD and primary blending source (closest to WD) simultaneously and plot them with filled and empty orange squares, respectively, in Figure \ref{blends}. When considering the photometry of the blending source along with the \emph{WISE} photometry, the resultant SEDs do not resemble that of an M dwarf or L/T brown dwarf, and are likely due to a background galaxy. The SWIRC data of the WD line up with the prediction for a hydrogen-dominated atmosphere for all but two of the blended targets, namely J1503 and J1552. These targets likely have an additional blending target not resolved in the SWIRC data that contributes to this deviation from the WD model.

Two targets (J0914 and J1530) have SEDs inconsistent with that of a debris disk and no neighboring point sources. While J0914 resembles a disk-hosting WD in that it has no NIR excess in conjunction with a MIR excess, the excess detected by \emph{WISE} is too strong to be reproduced by any of the geometrically thin, optically thick debris disk models that we fit it with. The \emph{WISE} excess is likely caused by a background red galaxy. The excess found at J1530, on the other hand, starts in the SDSS $z$-band which is uncharacteristic of a debris disk excess and enough to refute the presence of a disk.
%
%New Dusty WDs
\subsection{New Dusty WDs}\label{newdustywds}
A quarter of the WDs in our sample show an excess consistent with that of a debris disk and have no obvious source of contamination within the \emph{WISE} beam radius. The best fit disk parameters are listed in Table \ref{disks}. To evaluate the uncertainties in the disk temperatures and inclinations we perform a Monte Carlo analysis where we replace the observed photometric fluxes $f$ with $f + g \delta f$, where $\delta f$ is the error in flux and $g$ is a Gaussian deviate with zero mean and unit variance. For each of 7,000 sets of modified photometry, we repeat the analysis to derive best-fit parameters for the disk. We adopt the interquartile range of these parameters as the uncertainty. We find that the disk temperatures are uncertain by 200 K and the inclination by 20-25$^{\circ}$. The lack of data beyond 5 $\mu$m contributes to the uncertainty in the outer temperature of the disks.

The inner and outer radii are obtained from the inner and outer temperatures of the disk using Equation 1 of \cite{jura03}. The four disks all have similar widths ($\sim0.3 R_{\odot}$), however the radius of the inner edge of the disk orbiting J1537 is three times that of the others. The most massive (J1234) and the coolest (J1507) dusty WDs have the hottest and the coolest disk inner edges; 1900K and 800K, respectively. The temperature of the inner edge of the disk surrounding J1234 is hotter than the sublimation temperature of dust ($\sim 1500K$; \citealt{rafikov12}) and the flat disk model may need to be replaced with a warped disk in order to explain the SED (\citealt{jura09}).

J1507 is among the coolest dusty WDs known; WD 2115$-$560 (\citealt{vonhippel07,farihi09,jura09}) and G166-58 (\citealt{farihi08}) are the only previously known WD$+$disk systems with host WDs cooler than 10,000 $K$. It is also worth noting that J1234 is a 1.04 $M_{\odot}$ dusty WD with an estimated progenitor main-sequence mass of 5.4 Msol (\citealt{kalirai08,williams09}). J1234 is the most massive dusty WD currently known, and the discovery of its disk indicates that intermediate mass $\sim5~M_{\odot}$ main-sequence stars also host planets.
%
%The Frequency of Disks in the \emph{WISE} All Sky Survey
\subsection{The Frequency of Disks in the \emph{WISE} All Sky Survey}
Two WD$+$disk candidates identified by \cite{debes11}, J0110 and J0905, appear to be bonafide debris disks, due to the lack of an obvious contaminating source in the high resolution UKIDSS images, and must be included in the discussion of the frequency of disks in the WISE sample as a whole. Of the 12 WIRED disk candidates with UKIDSS data, only these two have no apparent contaminating source within a 6$\arcsec$ radius. With the inclusion of J0110 and J0905 there are now 12 confirmed debris disks in the \emph{WISE} sample. These include six previously known dusty WDs, four new dusty WDs presented by this work, and two dusty WDs with UKIDSS data. Out of 1527 WDs in the \emph{WISE} sample, 1020 are WD$+$M dwarf candidates and 42 are WD$+$brown dwarf candidates which leaves 465 single WDs observed by \emph{WISE}. Thus, the frequency of disks around single WDs in the \emph{WISE} sample is at least 2.6\% which is consistent with the 1-3\% disk frequency derived by \cite{farihi09}. One of the eight WD+disk candidates in our target list that remain unobserved (J0813) was recently revealed by \cite{wang13} to have a nearby blending source. The remaining seven disk candidates without NIR data could bring the WISE disk frequency as high as 4.1\%, consistent with the 4.3\% frequency found by \cite{barber12}.

If we restrict the frequency calculations to WDs with estimated W1 fluxes above the 74\% completeness limit of 50 $\mu$Jy, we could not include any of the four new disks in the calculation. The seven WIRED disk candidates included in the flux limited sample can, however, be pared down to four candidates because J1037 was misidentified as a WD, J2220 is a known WD$+$BD system, and J1559 has contaminated \emph{WISE} photometry. Therefore, the frequency of WD$+$disk systems in the flux limited WIRED sample is at most four out of 395 or 1\%. We calculate the confidence interval of the disk frequency in both the flux limited (0.7-1.8\%) and entire WISE sample (2.0-3.5\%) using the Bayesian method outlined in \cite{cameron11} and find that the two frequencies are consistent within a 2 $\sigma$ ($\sim 95\%$) confidence level.

None of the four new dusty WDs is identified by \cite{kleinman13} as having an atmosphere polluted with metals, but this is most likely due to the low resolution and low S/N of the SDSS spectroscopy data. A metal-free atmosphere does not preclude the existence of a debris disk (\citealt{hoard13}), however follow-up HIRES observations would be useful to confirm the state of pollution of these WD atmospheres. \cite{hoard13} report 7 WD$+$disk candidates, but 3 of them do not display pollution in their atmospheres. The warm inner disk temperatures in Table \ref{disks} imply that accretion is ongoing (\citealt{rafikov11a}) and we expect that high quality, high resolution spectra will reveal the presence of metals. Also, the four new dusty WDs in our sample should show $K$-band excess from the disks (see Figure \ref{disks}). $K$-band spectra would be useful to further characterize the source of the excess emission in these systems and to constrain the inner radii of the disks more precisely.
%
%
%
%CONCLUSIONS
\section{CONCLUSIONS}\label{conclusions}
We have confirmed the presence of dust surrounding four WDs with excess flux detected by \emph{WISE}. These four new dusty WDs, including the two bonafide UKIDSS$+$WISE disks, increases the total number of confirmed WD$+$disk systems from 29 to 35 (\citealt{farihi09,xu12,hoard13}). The discovery of these four new debris disks enriches the current dusty WD population with one of the coolest (J1507), hottest (J1537), and the most massive (J1234) WDs known to host circumstellar dust. Expanding the parameter space known to be hospitable to circumstellar dust at WD stars will not only guide future infrared excess searches, but also enhance our understanding of the formation and evolution of these remnant planetary systems.
%
%
%
%REFERENCES

%
%
%
%TABLES
%
%sample
\begin{deluxetable}{ccccccccc}\tablewidth{0pt}
\rotate
\tablecaption{SWIRC Photometry\label{sample}}
\tablehead{
\colhead{WIRED} & \colhead{$J$} & \colhead{$H$} & \colhead{$J_{blend}$} & \colhead{$H_{blend}$} & \colhead{Dist.} & \colhead{$T_{eff}$} & \colhead{$\log(g)$} & \colhead{Mass} \\ 
\colhead{} & \colhead{(mag)} & \colhead{(mag)} & \colhead{(mag)} & \colhead{(mag)} & \colhead{(pc)} & \colhead{($K$)} & \colhead{} & \colhead{($M_{\odot}$)} 
} 
\startdata
090611.00+414114.3 & 18.20$\pm$0.07 	& 18.27$\pm$0.11 	& 18.30$\pm$0.07 	& 17.29$\pm$0.11 	& 469 & 47637 & 7.91 & 0.65 \\
091411.11+434332.9 & 20.19$\pm$0.06 & 20.28$\pm$0.15 	&\nodata			&\nodata 			& 820 & 22621 & 7.85 & 0.56 \\
114758.61+283156.2 & 17.52$\pm$0.07 & 17.55$\pm$0.07 	&\nodata 			&\nodata 			& 134 & 12290 & 8.14 & 0.70 \\
122220.88+395553.9 & 19.15$\pm$0.02 & 18.93$\pm$0.02 	&\nodata			& 19.90$\pm$0.02 	& 350 & 11602 & 7.52 & 0.37 \\
123432.63+560643.0 & 18.30$\pm$0.04 	& 18.13$\pm$0.05 	&\nodata 			&\nodata 			& 133 & 13567 & 8.74 & 1.04 \\
131849.24+501320.6 & 20.16$\pm$0.07 & 20.17$\pm$0.08 	& 19.75$\pm$0.06 	& 18.97$\pm$0.06 	& 481 & 13305 & 7.96 & 0.59 \\
144823.67+444344.3 & 17.78$\pm$0.07 & 17.74$\pm$0.09 	& 20.77$\pm$0.09 	& 19.70$\pm$0.1 	& 267 & 18188 & 7.45 & 0.37 \\
150347.29+615847.4 & 19.97$\pm$0.05 & 19.41$\pm$0.05 	& 20.84$\pm$0.08 	& 19.94$\pm$0.07 	& 693 & 18006 & 7.86 & 0.53 \\
150701.98+324545.1 & 18.15$\pm$0.06 & 18.02$\pm$0.08 	&\nodata 			&\nodata 			& 114 & 7177 & 8.06 & 0.63 \\
151200.04+494009.7 & 19.44$\pm$0.07 & 19.37$\pm$0.07 	& 21.62$\pm$0.16 	& 19.98$\pm$0.09 	& 532 & 19527 & 7.71 & 0.49 \\
153017.00+470852.4 & 19.15$\pm$0.05 & 18.87$\pm$0.05 	&\nodata			&\nodata 			& 703 & 15479 & 7.53 & 0.39 \\
153725.71+515126.9 & 17.75$\pm$0.05 & 17.54$\pm$0.07 	&\nodata 			&\nodata 			& 199 & 24926 & 8.20 & 0.73 \\
154038.67+450710.0 & 18.36$\pm$0.06 & 18.23$\pm$0.08 	& 19.93$\pm$0.07 	& 19.03$\pm$0.04 	& 154 & 8824 & 8.03 & 0.62 \\
155206.11+391817.2 & 20.56$\pm$0.09 & 19.72$\pm$0.10	& 21.03$\pm$0.10 	& 19.88$\pm$0.1 	& 945 & 20040 & 7.80 & 0.53 \\
155955.27+263519.2 & 16.44$\pm$0.05 & 16.42$\pm$0.06 	& 20.52$\pm$0.09 	& 19.54$\pm$0.09 	& 67 & 11890 & 8.28 & 0.78 \\
165747.02+624417.4 & 19.21$\pm$0.07 & 19.22$\pm$0.10 	& 20.96$\pm$0.09 	& 19.93$\pm$0.11 	& 351 & 14241 & 7.87 & 0.55 \\
\enddata
\end{deluxetable}\nopagebreak
% 
%disks
\begin{deluxetable}{cccccc}\tablewidth{0pt}
\tablecaption{Best Fit Disk Parameters\label{disks}}
\tablehead{\colhead{WIRED} & \colhead{$T_{in}$} & \colhead{$T_{out}$} & \colhead{Incl.} & \colhead{$R_{in}$} & \colhead{$R_{out}$} \\ 
\colhead{} & \colhead{($K$)} & \colhead{($K$)} & \colhead{($\degree$)} & \colhead{($R_{\odot}$)} & \colhead{($R_{\odot}$)} } 
\startdata
114758.61+283156.2 & 1400 & 600 & 60 & 0.126 &0.389  \\
123432.63+560643.0 & 1900 & 500 & 70 & 0.062  & 0.367  \\
150701.98+324545.1 & 800 & 300 & 0 &  0.134& 0.494  \\
153725.71+515126.9 & 1200 & 800 & 30 &  0.414&0.711  
\enddata\tablecomments{Typical uncertainties for disk inner and outer temperatures are 200K and  20-25\% for inclination.}
\end{deluxetable}\clearpage
%
%
%
%FIGURES
%
%blends
\begin{figure}[h!]
	\centering
	\includegraphics[width=\linewidth,angle=0]{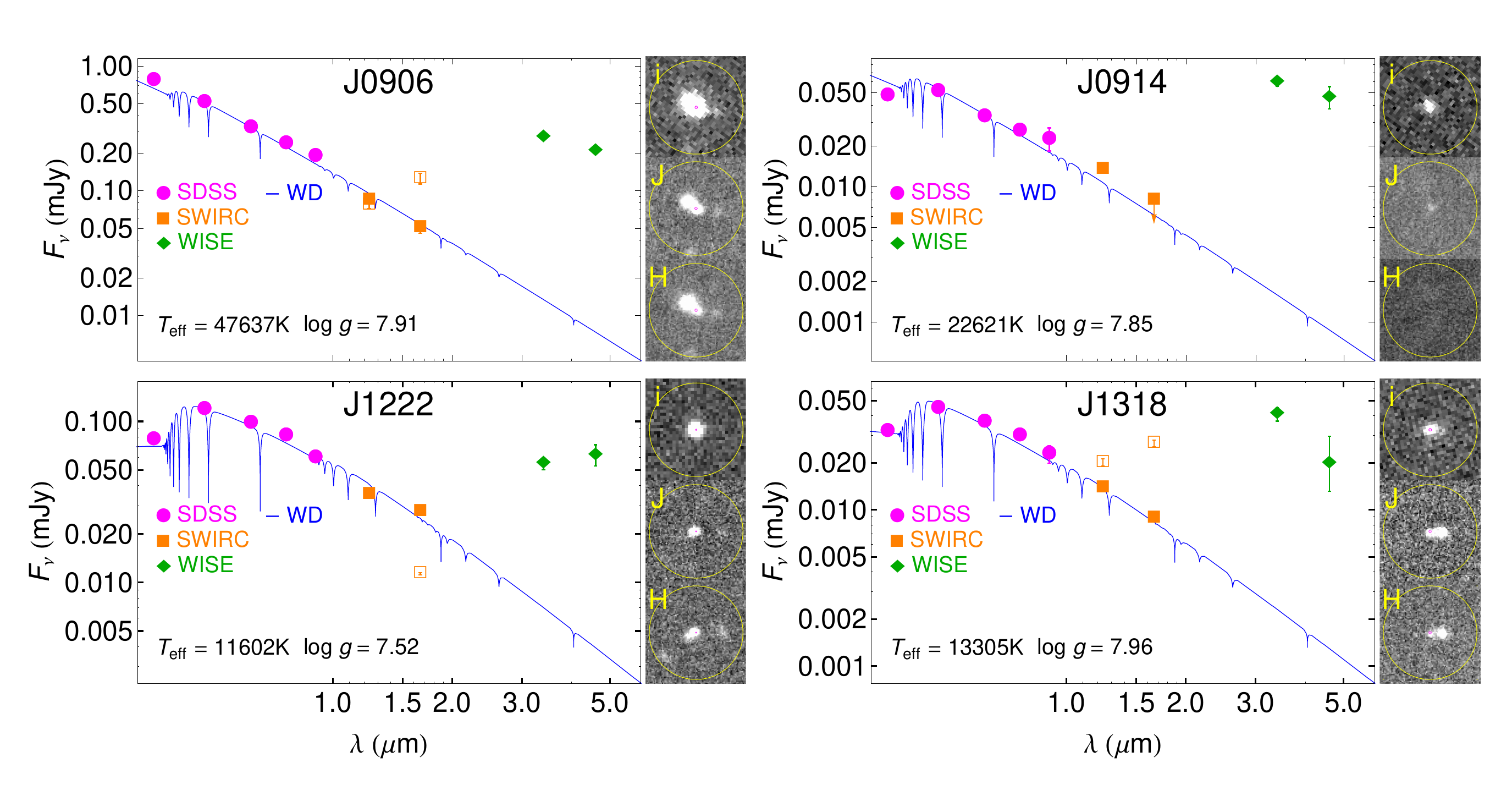}
\caption{The spectral energy distribution of WDs with nearby point sources. Pure hydrogen atmosphere models (\citealt{tremblay09,tremblay10,gianninas11}), shown in blue, are normalized to the SDSS $griz$-bands. SDSS, SWIRC, and \emph{WISE} photometry are shown as magenta circles, orange squares (empty squares for blended point source), and green diamonds, respectively. ALLWISE data are plotted where available. The SDSS $i$-band and SWIRC $J$- and $H$-band science images of the corresponding WD field are shown to the right. The yellow circle indicates the 6$\arcsec$ \emph{WISE} beam. The small magenta circle is centered on the WD's SDSS J2000 coordinates with a radius equal to the proper motion since the SDSS DR7 data were taken and is used to identify the WD. Extraneous objects within the yellow circle are likely the source of the excesses detected in the \emph{WISE} data.\label{blends}}
\end{figure}
\begin{figure}[h!]
	\includegraphics[width=\linewidth]{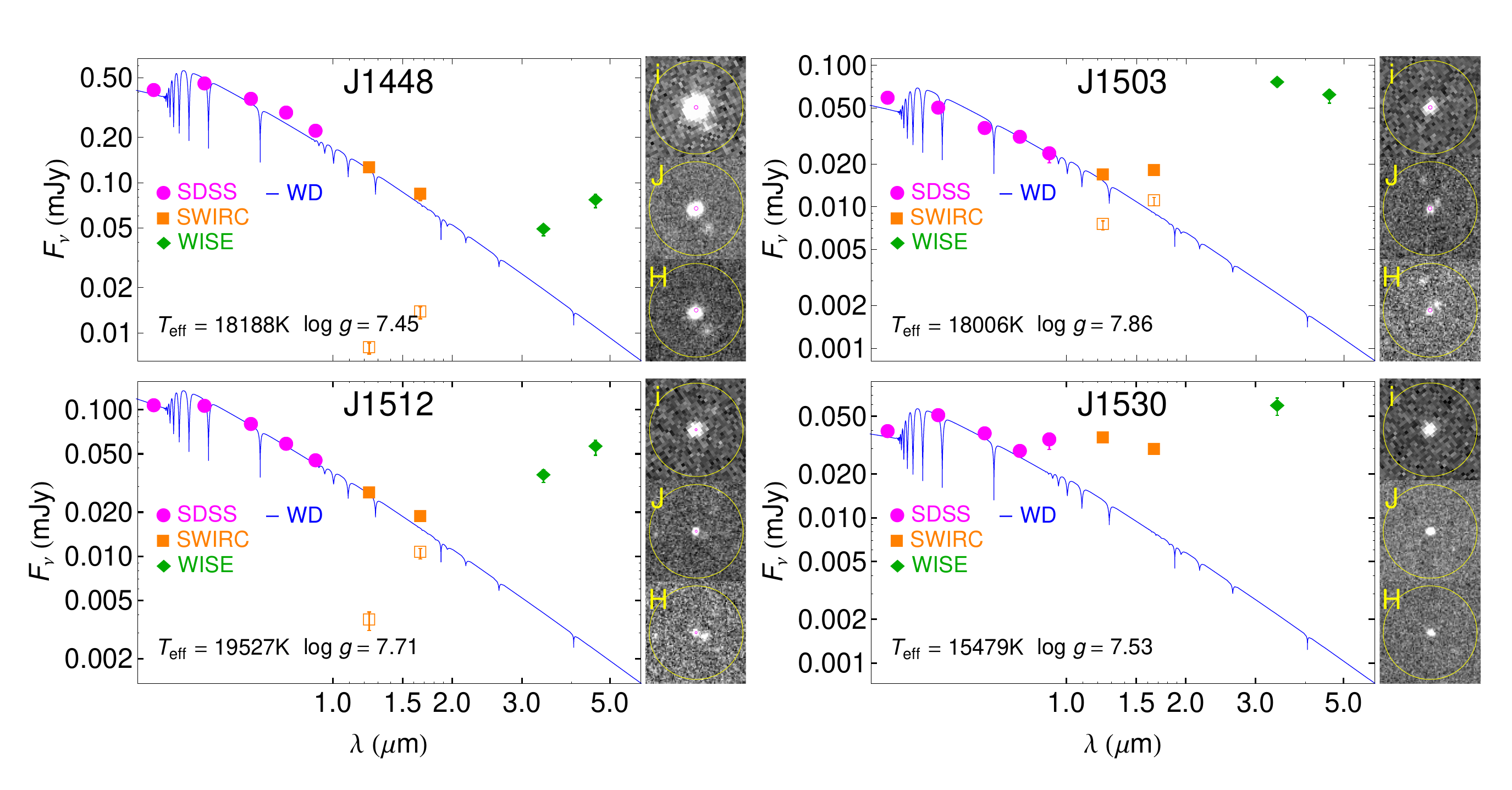}
	\figurenum{1}
	\caption{Continued.}
\end{figure}
\begin{figure}[h!]
	\includegraphics[width=\linewidth]{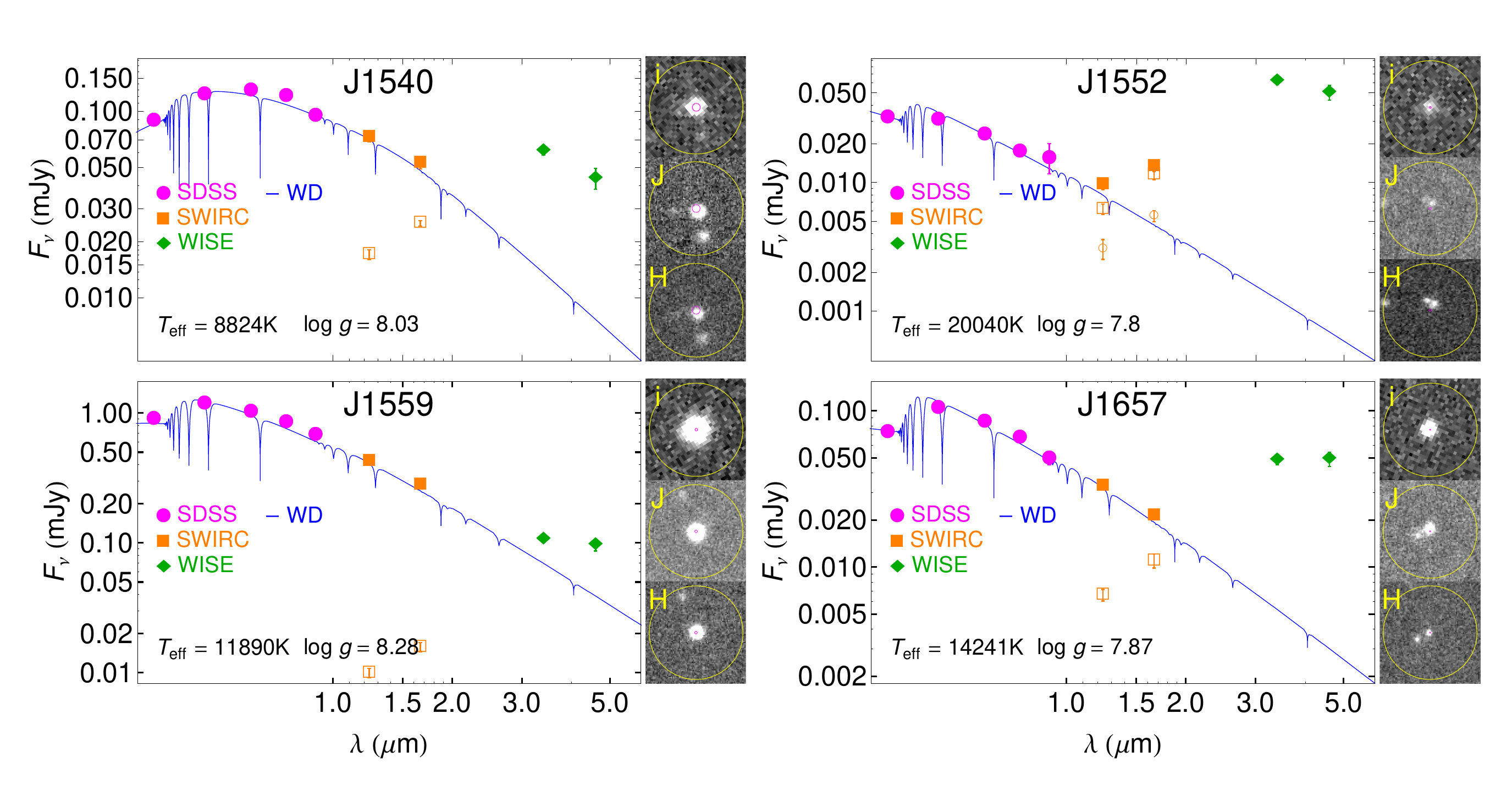}
	\figurenum{1}
	\caption{Continued.}
\end{figure}
%
%blendcolors
%
%dusty
\begin{figure}[h!]
	\centering
	\includegraphics[width=\linewidth]{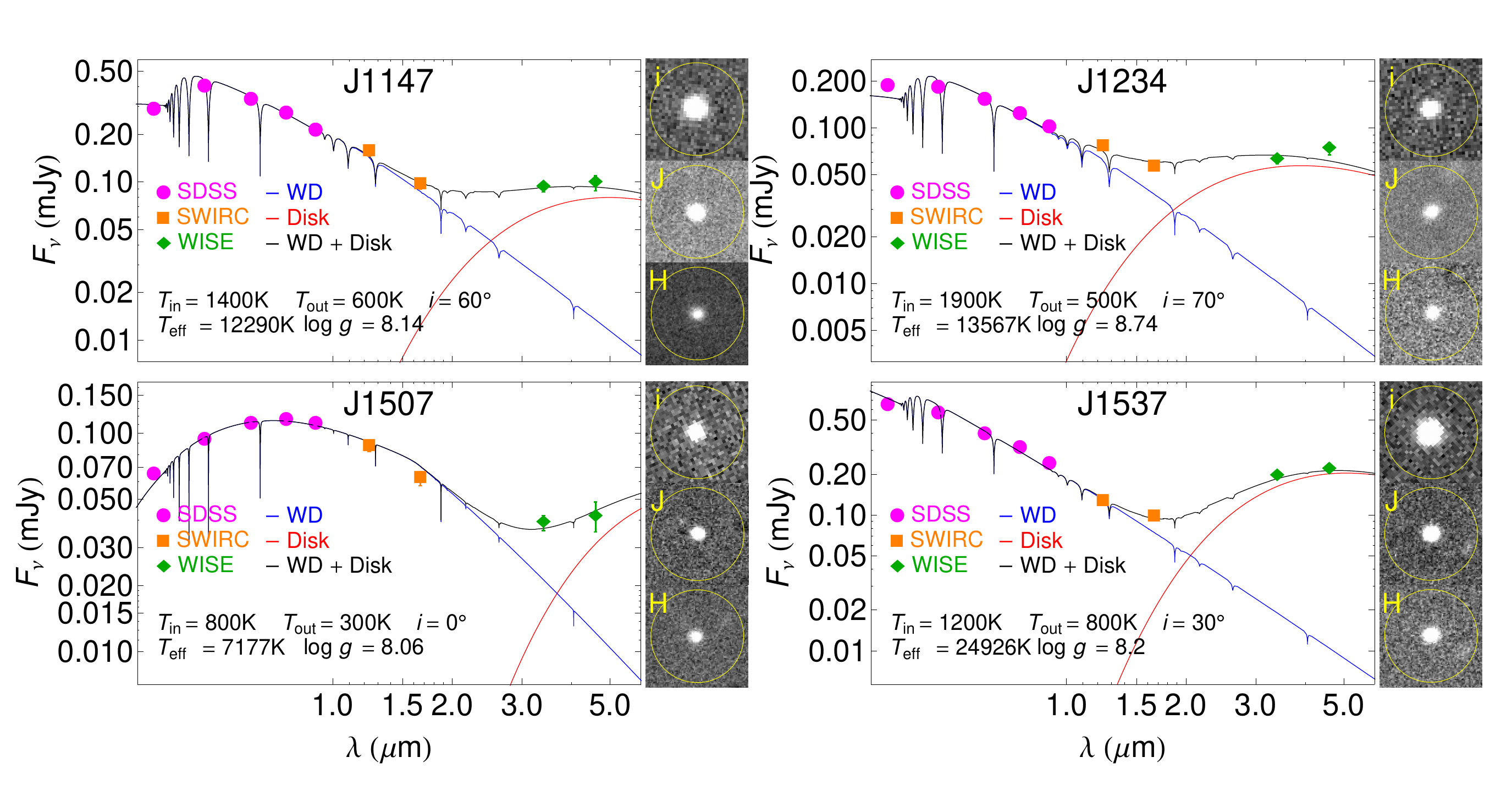}
	\caption{The spectral energy distribution of the four isolated WDs in our sample. The SED components are the same as those described in Figure \ref{blends} except for the disk model, which is shown in red and the WD$+$disk combination which is shown in black here. The science images of these WDs show no obvious sign of contamination that could explain the excesses found in the \emph{WISE} data. These excesses are well reproduced by geometrically flat, optically thick disk models (\citealt{jura03}). \label{dusty}}
\end{figure}
%
%diskcolors
\end{document}